\documentclass{article}
\usepackage[T1]{fontenc} 
\usepackage[utf8]{inputenc} 
\usepackage{ismir,amsmath,cite,url}
\usepackage{graphicx}
\usepackage{color}
\usepackage{enumitem}

\usepackage{lineno}

\title{Controllable deep melody generation \\ via hierarchical music structure representation}

\fourauthors
  {Shuqi Dai} {Carnegie Mellon university \\ {\tt shuqid@cs.cmu.edu}}
  {Zeyu Jin} {Adobe Inc. \\ {\tt zejin@adobe.com}}
  {Celso Gomes} {Adobe Inc. \\ {\tt cegomes@adobe.com}}
  {Roger B. Dannenberg} {Carnegie Mellon University \\ {\tt rbd@cs.cmu.edu}}

\def\authorname{S. Dai, Z. Jin, C. Gomes and R. Dannenberg}

\begin{document}

\maketitle
\begin{abstract}

Recent advances in deep learning have expanded possibilities to generate music,
but generating a customizable full piece of music with consistent long-term structure remains a challenge. This paper introduces \textit{MusicFrameworks}, a hierarchical music structure representation and a multi-step generative process to create a full-length melody guided by long-term repetitive structure, chord, melodic contour, and rhythm constraints. We first organize the full melody with section and phrase-level structure. To generate melody in each phrase, we generate rhythm and basic melody using two separate transformer-based networks, and then 
generate the melody conditioned on the basic melody, rhythm and chords in an auto-regressive manner. 
By factoring music generation into sub-problems, our approach allows simpler models and requires less data.
To customize or add variety, one can alter chords, basic melody, and rhythm structure in the music frameworks, letting our networks generate the melody accordingly. 
Additionally, we introduce new features to encode musical positional information, rhythm patterns, and melodic contours based on musical domain knowledge. 
A listening test reveals that melodies generated by our method are rated as good as or better than human-composed music in the POP909 dataset about half the time. 

\end{abstract}

\section{Introduction}\label{sec:introduction}
Music generation is an important component of computational and AI creativity, leading to many potential applications including automatic background music generation for video,
music improvisation in human-computer music performance 
and customizing stylistic music for individual music therapy,
to name a few. While works such as MelNet~\cite{vasquez2019melnet} and JukeBox~\cite{dhariwal2020jukebox} have demonstrated a degree of success in generating music in the audio domain, the majority of the work is in the symbolic domain, i.e., the score, as this is the most fundamental representation of music composition. 
Research has tackled this question from many angles, including monophonic melody generation \cite{medeot2018structurenet}, polyphonic performance generation~\cite{huang2018music} and drum pattern generation\cite{wei2019generating}. This paper focuses on melody generation, a crucial component in music writing practice. 

Recently, deep learning has demonstrated success in capturing implicit rules about music from the data, compared to conventional rule-based and statistical methods~\cite{huang2018music, huang2020pop, hakimi2020bebopnet}.
However, there are three problems that are difficult to address: (1) Modeling larger scale music structure and multiple levels of repetition
as seen in popular songs,
(2) Controllability 
to match music to video or create desired tempo, styles, and mood,
and (3) Scarcity of training 
data due to limited curated and machine-readable compositions, especially in a given style.
Since humans can imitate music styles with just a few samples, there is reason to believe there exists a solution that enables music generation with few samples as well. 

We aim to explore automatic melody generation with multiple levels of structure awareness and controllability. Our focus is on (1) addressing structural consistency inside a phrase and on the global scale, and (2) giving explicit control to users to manipulate melody contour and rhythm structure directly.
Our solution, \textit{MusicFrameworks}, is based on the design of hierarchical music representations we call \textit{music frameworks} inspired by Hiller and Ames \cite{10.2307/832731}.
A music framework is an abstract hierarchical description of a song, including high-level music structure such as repeated sections and phrases, and lower-level representations such as rhythm structure and melodic contour. 
The idea is to represent a piece of music by music frameworks, and then learn to generate melodies from music frameworks.
Controllability is achieved by editing the music frameworks at any level (song, section and phrase); we also present methods that generate these representations from scratch. \textit{MusicFrameworks} can create long-term music structures, including repetition, by factoring music generation into sub-problems, allowing simpler models and requiring less data.

Evaluations of the \textit{MusicFrameworks} approach include objective measures to show expected behavior and subjective assessments. We compare human-composed melodies and melodies generated under various conditions to study the effectiveness of music frameworks.  We summarize our contributions as follows:
(1) devising a hierarchical music structure representation and approach called {\it MusicFrameworks} capable of capturing repetitive structure at multiple levels, 
(2) enabling controllability at multiple levels of abstraction through music frameworks, 
(3) a set of methods that analyze a song to derive music frameworks that can be used in music imitation and subsequent deep learning processes,  
(4) a set of neural networks that generate a song using the \textit{MusicFrameworks} approach,
(5) useful musical features and encodings to introduce musical inductive biases into deep learning,
(6) comparison of different deep learning architectures for relatively small amounts of training data and a sizable listening test evaluating the musicality of our method against human-composed music.

\section{Related Work}\label{sec:relatedwork}

Automation of music composition with computers can be traced back to 1957 \cite{Hiller}.  Long before representation learning,
musicians looked for models that explain the generative process of music\cite{10.2307/832731}. Early music generation systems often relied on generative rules or constraint satisfaction~\cite{10.2307/832731, cope1991computers, copealgo, copecom}. Subsequent approaches replaced human learning of rules with machine learning, such as statistical models \cite{schulze2011music} and connectionist approaches \cite{boulanger2012modeling}. Now, deep learning has emerged as one of the most powerful tools to encode implicit rules from data \cite{briot2019deep, liang2016bachbot, hadjeres2017deepbach, huang2019counterpoint, huang2018music}. 

One challenge of music modelling is capturing repetitive patterns and long-term dependencies. There are a few models using rule-based and statistical methods to construct long-term repetitive structure in classical music \cite{collins2017computer} and pop music \cite{elowsson2012algorithmic, dai2021personalized}. Machine learning models with memory and the ability to associate context have also been popular in this area and include LSTMs and Transformers ~\cite{vaswani2017attention, huang2018music, musenet, huang2020pop}, which operate by generating music one or a few notes at a time, based on information from previously generated notes. These models enable free generation and motif continuation, but it is difficult to control the generated content. StructureNet \cite{medeot2018structurenet}, PopMNet \cite{popmnet2020} and Racchmaninof \cite{collins2017computer} are more closely related to our work in that they introduce explicit models for music structure.

Another thread of work enables a degree of controllability by modeling the distribution of music via an intermediate representation (embedding). One such approach is to use Generative Adversarial Networks (GANs) to model the distribution of music ~\cite{Goodfellow2014GenerativeAN, yu2017seqgan, Dong_Hsiao_Yang_Yang_2018}. GANs learn a mapping from a point $z$ sampled from a prior distribution to an instance of generated music $x$ and hence represents the distribution of music with $z$. Another method is the Autoencoder, consisting of an encoder transforming music $x$ into embedding $z$ and a decoder that reconstructs music $x$ again from embedding $z$. The most popular models are  Variational Auto-Encoders (VAE) and their variants \cite{kingma2013auto, 47078, yang2019deep, Tan2020MusicFC, kawaiattributes, Wang2020LearningIR}. These models can be  controlled by manipulating the embedding, for example, mix-and-matching embeddings of different pitch contours and rhythms \cite{47078, yang2019deep, chen2020music}. However, a high-dimensional continuous vector has limited interpretability and thus is difficult for a user to control; it is also difficult to model full-length music with a simple fixed-length representation. In contrast, our approach uses a hierarchical music representation (i.e., {\it music framework}) as an ``embedding'' of music that encodes long-term dependency in a form that is both interpretable and controllable. 

\section{Method}\label{sec:method}
\begin{figure}[tb]
 \centerline{
 \includegraphics[width=0.98\columnwidth]{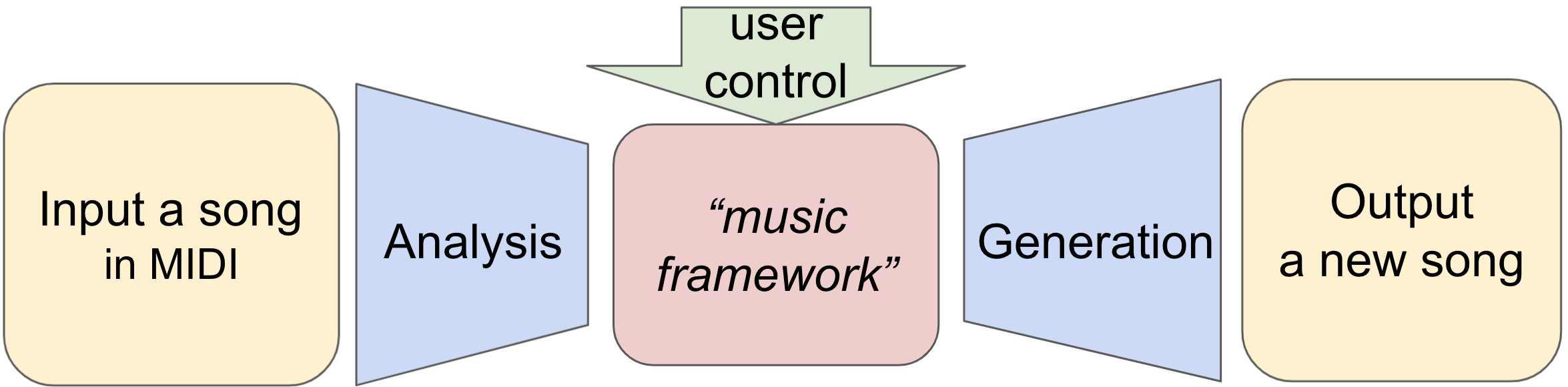}}
 \caption{Architecture of \textit{MusicFrameworks}.}
 \label{fig:architecture}
\end{figure}

We describe a controllable melody generation system that uses hierarchical music representation to generate full-length pop song melodies with multi-level repetition and structure. 
As shown in Figure \ref{fig:architecture}, a song is input in MIDI format. We analyze its \textit{music framework}, which is an abstracted description of the music ideas of the input song. Then we use the music framework to generate a new song with deep learning models. 

Our work is with pop music because structures are relatively simple and listeners are generally familiar with the style and thus able to evaluate compositions. We use a Chinese pop song dataset, POP909 \cite{pop909-ismir2020}, and use its cleaned version \cite{dai2020automatic} (with more labeling and corrections) for training and testing in this paper. 
We further transpose all the major songs' key signatures into C major. We use integers 1--15 to represent scale degrees in C3--C5, and 0 to represent a rest. For rhythm, we use the $16^{th}$ note as the minimum unit. A note in the melody is represented as $(p, d)$, where $p$ is the pitch number from 0 to 15, and $d$ is 
duration in sixteenths.
For chord progressions, we use integers 1--7 to represent seven scale degree chords in the major mode. (We currently work only with triads, and convert seventh chords into corresponding triads).

\subsection{Music Frameworks Analysis}\label{sec:analysis}
\begin{figure}[tb]
 \centerline{
 \includegraphics[width=0.95\columnwidth]{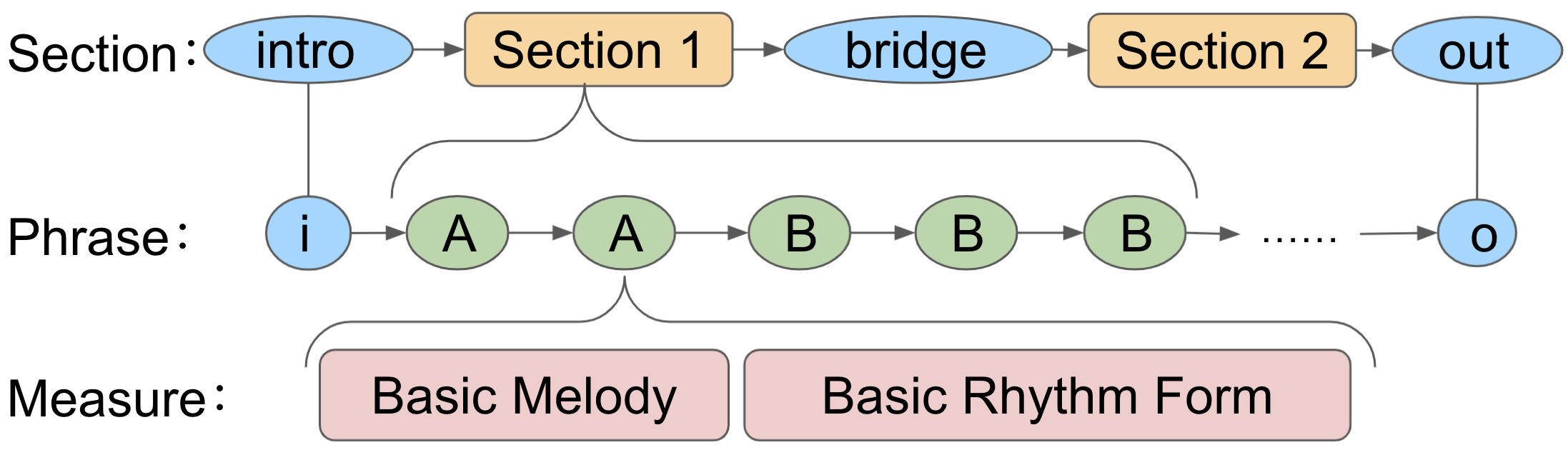}}
 \caption{An example music framework.}
 \label{fig:musicframework}
\end{figure}

\begin{figure}[tb]
 \centerline{
 \includegraphics[width=0.95\columnwidth]{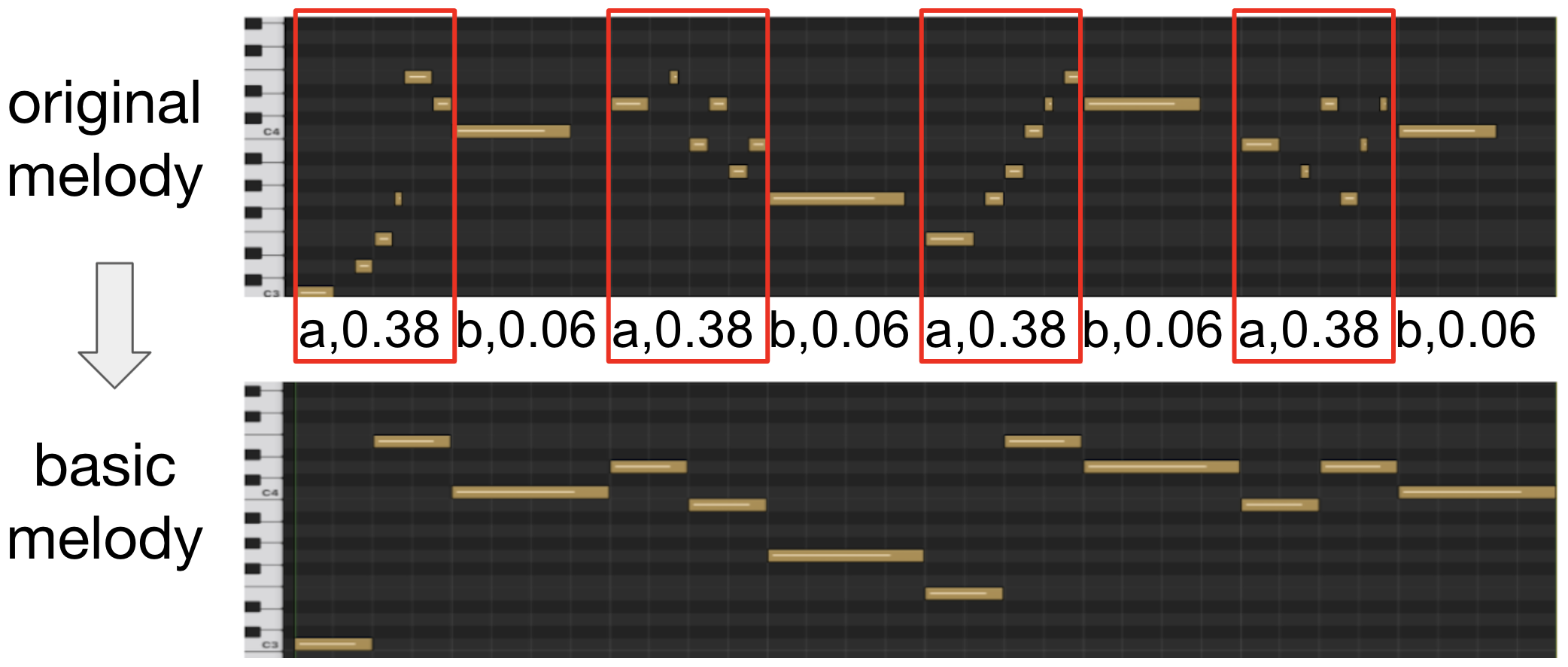}}
 \caption{An example melody and its basic melody. The basic rhythm form also appears below the original melody as ``a,0.38 b,0.06, ...'' indicating similarity and complexity. }
 \label{fig:basic_components}
\end{figure}

As shown in Figure \ref{fig:musicframework}, a music framework contains two parts: (1) section and phrase-level structure analysis results; (2) basic melody and basic rhythm form within each phrase. A phrase is a small-scale segment that usually ranges from 4 to 16 measures. Phrases can be repeated as in AABBB shown in Figure \ref{fig:musicframework}. 
Sections contain multiple phrases, e.g., the illustrated song has an intro, a main theme section (phrase A as verse and phrase B as chorus), a bridge section followed by a repeat of the theme, and an outro section, which is a typical pop song structure.  We extract the section and phrase structure based on finding approximate repetitions following the work of \cite{dai2020automatic}.

Within each phrase, the \textit{basic melody} is an abstraction of melody and contour. Basic melody is a sequence of half notes representing the most common pitch in each 2-beat segment of the original phrase (see Figure \ref{fig:basic_components}).
The \textit{basic rhythm form} consists of a per-measure descriptor with two components: a pattern label based on a rhythm similarity measure \cite{dai2020automatic} (measures with matching labels are similar) and a numerical rhythmic complexity, which is simply the number of notes divided by 16.

With the analysis algorithm, we can process a music dataset such as POP909 for subsequent machine learning and music generation. The music frameworks enable controllability via examples in which a user can also mix and match different music frameworks from multiple songs. For example, a new song can be generated using the structure from song A, basic melody from song B, and basic rhythm form from song C. Users can also edit or directly create a music framework for even greater control. Alternatively, we also created generative algorithms to create new music frameworks without any user intervention as described in subsequent sections.

\begin{figure}[tb]
 \centerline{
 \includegraphics[width=0.95\columnwidth]{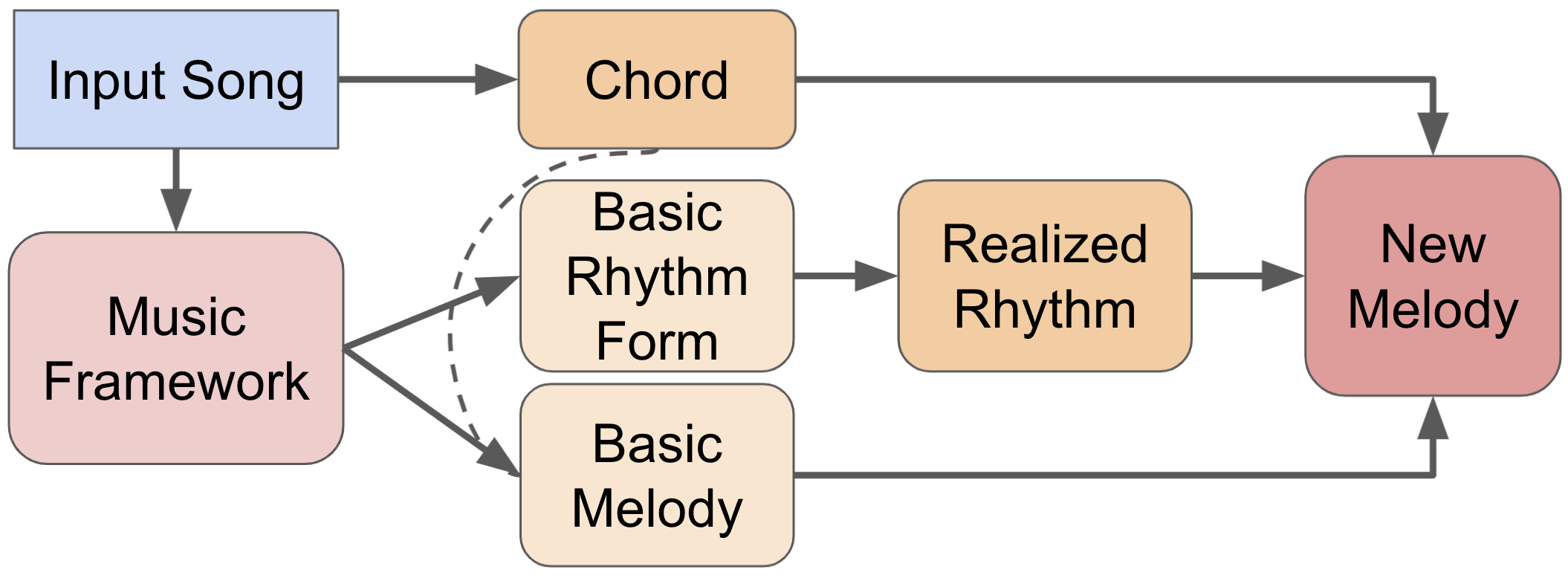}}
 \caption{Generation process from music frameworks within each phrase.}
 \label{fig:generation}
 \vspace{-1em}
\end{figure}

\subsection{Generation Using Music Frameworks}

At the top level, section and phrase structure can be provided by a user or simply selected from a library of already analyzed data. We considered several options for imitation at this top level: (1) copy the first several measures of melody from the previously generated phrase (teacher forcing mode) and then complete the current phrase; (2) use the same or similar basic melody from the previous phrase to generate an altered melody with a similar melodic contour; (3) use the same or similar basic rhythm form of the previous phrase to generate a similar rhythm. These options leave room for users to customize their personal preferences. In this study, we forgo all human control by randomly choosing between the first and second option.

At the phrase level, as shown in Figure \ref{fig:generation}, we first generate a basic melody (or a human provides one). 
Next, we
generate rhythm using the basic rhythm form. Finally, we generate a new melody given the basic melody, generated rhythm, and chords copied from the input song.

\subsubsection{Basic Melody Generation}\label{sec:basicmelody}
We use an auto-regressive approach to generate basic melodies.
The input $x_i = (\text{pos}_i, c_i, ...)$ ($i \in {1,...,n}$) is a set of features that guides basic melody generation where $\text{pos}_i$ is the positional feature of the $i^{th}$ note and $c_i$ represents contextual chord information (neighboring chords). We denote $p_i$ of the $i^{th}$ note. Here we fix the duration of each note in the basic melody to the half-note as in the analysis algorithm described in Section \ref{sec:analysis}. $c_i$ contains the previous, current and next chords and their lengths for the $i^{th}$ note. $pos_i$ includes the position of the $i^{th}$ note in the current phrase and a long-term positional feature indicating whether the current phrase is at the end of a section or not. 

\subsubsection{Network Architecture}
\begin{figure}[tb]
 \centerline{
 \includegraphics[width=1.0\columnwidth]{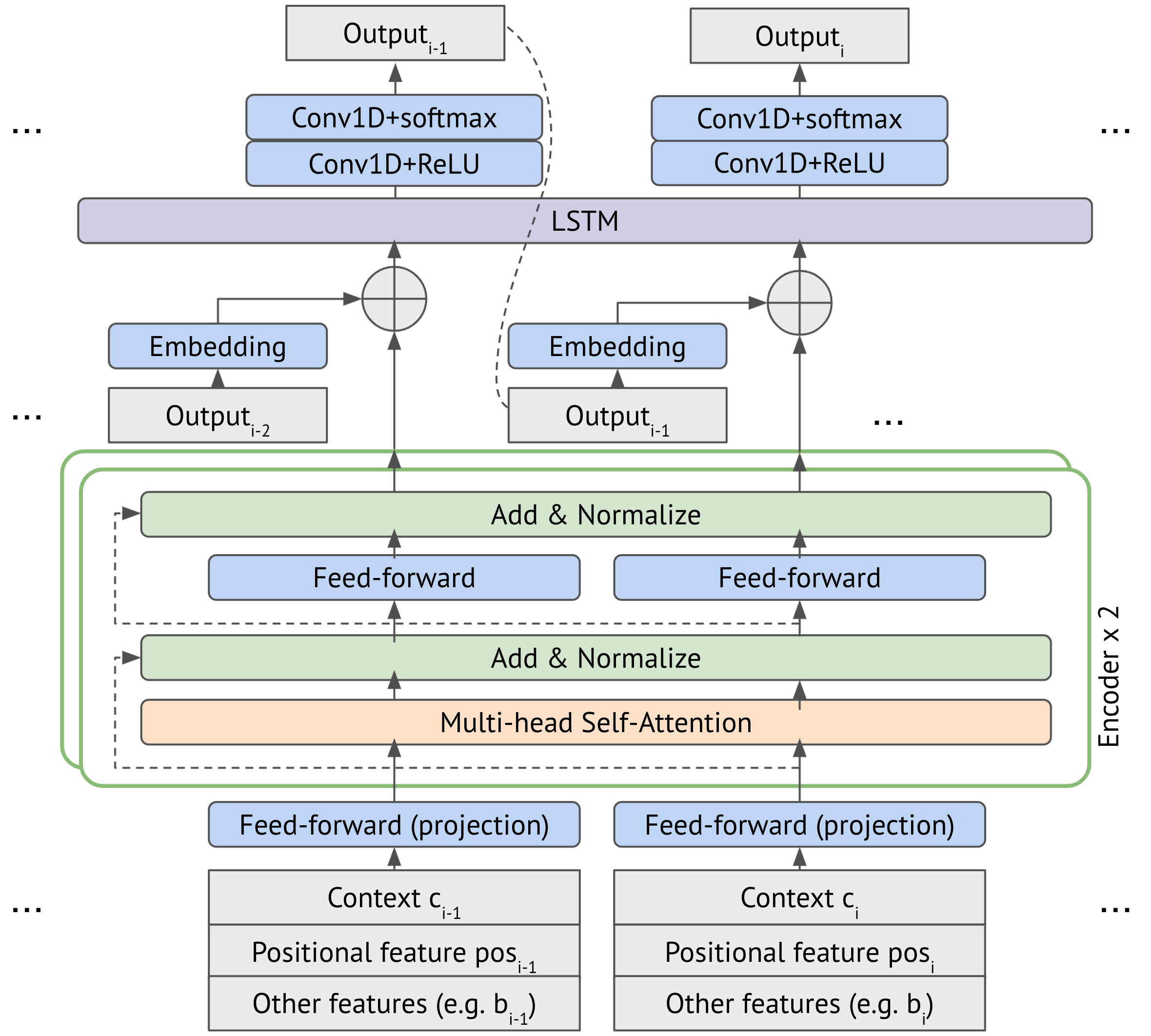}}
 \caption{Transformer-LSTM architecture for melody, basic melody and rhythmic pattern generation.}
 \label{fig:transformer-lstm}
 \vspace{-1em}
\end{figure}

We use an auto-regressive model based on Transformer and LSTM. The architecture (Figure \ref{fig:transformer-lstm}) consists of an encoder and a decoder. The encoder has two layers of transformers that learn a feature representation of the inputs (e.g. positional encodings and chords). The decoder concatenates the encoded representation and the last predicted note as input and passes them through one unidirectional LSTM followed by two layers of $1D$ convolutions of kernel size 1. Both the input and the last predicted notes to the decoder are passed through a projection layer (aka. a dense layer) respectively before they are processed by the network. The final output is the next note predicted by the decoder via categorical distribution $Pr(p_i | X, p_1,...,p_{i-1})$. 
We also tried using other deep neural network architectures such as a pre-trained full Transformer with random masking (described in Section \ref{sec:objexperiment}) for comparison.

\subsubsection{Sampling with Dynamic Time Warping Control}
In the sampling stage, we tried three ways to autoregressively generate the basic melody sequence: (1) randomly sample from the estimated posterior distribution of $p_i$ at each step;
(2) randomly sample 100 generated sequences and pick the one with highest overall estimated probability; 
(3) beam search sampling according to the estimated probability. Apart from the above three sampling methods, we also want to control the basic melody contour shape in order to generate similar or repeated phrases.  We use a melody contour rating function (based on Dynamic Time Warping) \cite{dai2021personalized} to estimate the contour similarity between two basic melodies. When we want to generate a repetition phrase that has a similar basic melody compared to a previous phrase, we estimate the contour similarity rating between the generated basic melody and the reference basic melody. We only accept basic melodies with a similarity above a threshold of $0.7$.

\subsubsection{Realized Rhythm Generation}
We now turn to the lower level of music generation that transforms music frameworks into realized songs. The first step is to determine the note onsets, namely the rhythm.
Instead of generating note onset time one by one, we generate 2-beat rhythm patterns, which more readily encode rhythmic patterns and metrical structure.
It is also easier to model (and apparently to learn) similarity using rhythm patterns than with sequences of individual durations. 

We generate 2-beat patterns sequentially under the control of a basic rhythm form. There are 256 possible rhythm patterns with a 2-beat length using our smallest subdivision of sixteenths. For each rhythm pattern $r_i$, the input of the rhythm generation model is $x_i = (r_{i-1}, \text{brf}_{i}, \text{pos}_{i})$, where $r_{i-1}$ is the previously generated rhythm pattern, $\text{brf}_{i}$ is the index of the first measure similar to it (or the current measure if there is no previous reference) and the current measure complexity;  $\text{pos}_{i}$ contains three positional components: (1) the position of the $i^{th}$ pattern in the current phrase; (2) a long-term positional feature indicating whether the current phrase is at the end of a section or not; (3) whether the $i^{th}$ rhythm pattern starts at the barline or not. We also use a Transformer-LSTM architecture (Figure \ref{fig:transformer-lstm}), but with different model settings (size). In the sampling stage, we use beam search.

\subsubsection{Realized Melody Generation}
We generate melody from a basic melody, a rhythm and a chord progression using another Transformer-LSTM architecture similar to generating basic melody in Figure \ref{fig:transformer-lstm}. 
In this case, the index $i$ represents the $i^{th}$ note determined by the rhythm. The input feature $x_i$ also includes the current note's duration, the current chord,
the basic melody pitch, and three positional features for multiple-level structure guidance: the two positional features for basic melody generation (Section \ref{sec:basicmelody}) and the offset of the current note within the current measure. We also experimented with other deep neural network architectures described in Section \ref{sec:objexperiment} for comparison. To sample a good sounding melody, we randomly generate 100 sequences by sampling the autoregressive model. We pick the one with the highest overall estimated probability. More details about the network are in Section \ref{sec:objexperiment}. 

\section{Experiment and Evaluation}\label{sec:experiment}

\subsection{Model Evaluation and Comparison}\label{sec:objexperiment}
As a model-selection study, we compared the ability of different deep neural network architectures implementing {\it MusicFrameworks} to predict the next element in the sequence. \textit{Basic Melody Accuracy} is the percent of correct predictions of the next pitch of the basic melody (half notes). \textit{Rhythm Accuracy} is the percent of correctly predicted 2-beat rhythm patterns. \textit{Melody Accuracy} is the accuracy of next pitch prediction.

We used 4188 phrases from 528 songs in major mode from the POP909 dataset, using 90\% of them as training data and the other 10\% for validation. The first line in Table \ref{tab:accuracy} represents the Transformer-LSTM models introduced in Section \ref{sec:method}. In all three networks, the projection size and feed forward channels are 128; there are 8 heads in the multi-head encoder attention layer; LSTM hidden size is 64; dropout rate for basic melody and realized melody generation is 0.2, dropout rate for rhythm generation is 0.1; decoder input projection size is 8 for rhythm generation and 17 for others. For learning rate, we used the Adam optimizer with $\beta 1 = 0.9, \beta 2 = 0.99, \epsilon = 10^{-6}$, and the same formula in \cite{vaswani2017attention} to vary the learning rate over the course of training, with 2000 warmup steps.

We compared this model with several alternatives: the second model is a bi-directional LSTM followed by a uni-directional LSTM (model size is 64 in both). The third model is a Transformer with two layers of encoder and two layers of decoder (with same parameter settings as Transformer-LSTM), and we first pre-trained the encoder with 10\% of random masking of input (similar to training in BERT \cite{Devlin2019BERTPO}), and then trained the encoder and decoder together. No music frameworks (the fourth line) means generate without basic melody or basic rhythm form, using a Transformer-LSTM model. The results in Table \ref{tab:accuracy} show that the Transformer-LSTM achieved the best accuracy. The full Transformer model performed poorly on this relatively small dataset due to overfitting. Also, in both rhythm and melody generation, the \textit{MusicFrameworks} approach significantly improves the model accuracy.

\begin{table}
 \begin{center}
 \begin{tabular}{|l|l|l|l|}
  \hline
   & Basic Melody & Rhythm & Melody  \\ 
  \hline
 Trans-LSTM  & 38.7\%       &   \textbf{50.1\%}        &      \textbf{55.2\%}                         \\
  \hline
 LSTM &   \textbf{39.8\%}     & 43.6\%   &     51.2\%        \\ 
\hline
 Transformer &  30.9\%      & 25.8\%   &    39.3\%   \\ 
\hline
No M.F. &  NA    &   33.1\%   & 37.4\%\\
\hline
 \end{tabular}
\end{center}
\vspace{-0.1in}
 \caption{Validation Accuracy of different model architectures. ``No M.F.'' means no music frameworks used here.}
 \label{tab:accuracy}
  \vspace{-1em}
\end{table}

\subsection{Objective Evaluation}

\begin{figure}[t]
 \centerline{
 \includegraphics[width=0.95\columnwidth]{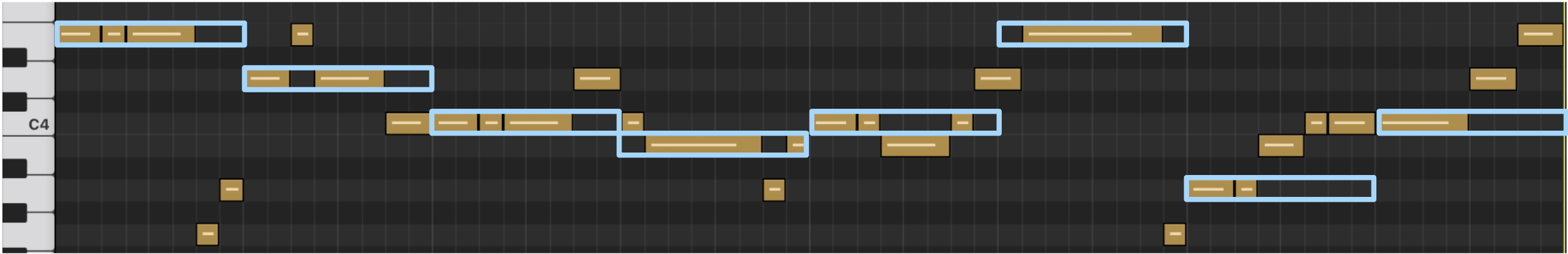}}
 \vspace{-0.1in}
 \caption{This is a generated melody (yellow piano roll) from our system following the input basic melody (blue frame piano roll).}
 \label{fig:generated_bm}
\end{figure}

\begin{figure}[t]
 \centerline{
 \includegraphics[width=0.95\columnwidth]{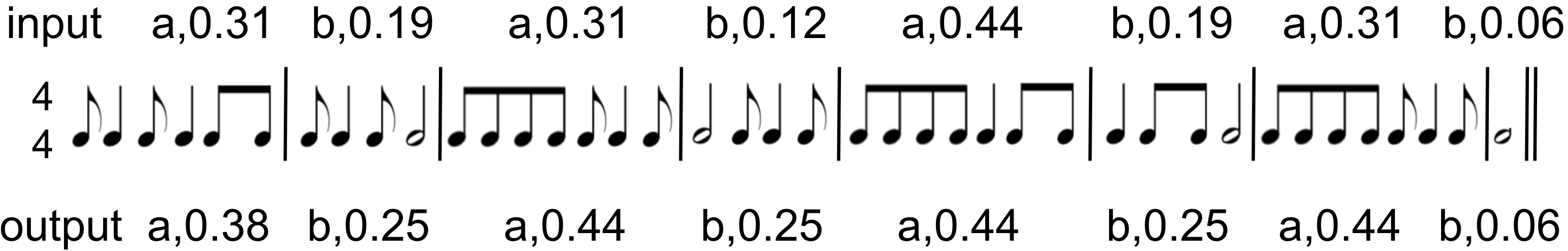}}
 \vspace{-0.1in}
 \caption{A generated rhythm from our system given the input basic rhythm form. The analyzed basic rhythm form of the output is very similar to the input.}
 \label{fig:generated_rhythm}
 \vspace{-1em}
\end{figure}

We use Transformer-LSTM model for all further evaluations. First, we examine whether music frameworks promote controllability. We aim to show that given a basic melody and rhythm form as guidance, the model can generate a new melody that follows the contour of the basic melody, and has a similar rhythm form (Figure \ref{fig:generated_bm} and \ref{fig:generated_rhythm}).

For this ``sanity check,'' we randomly picked 20 test songs and generated 500 alternative basic melodies and rhythm forms. After generating and analyzing 500 phrases, we found the analyzed basic melodies match the target (input) basic melodies with an accuracy of 92.27\%; the accuracy of rhythm similarity labels is 94.63\%; the rhythmic complexity matches the target (within $0.2$) 81.79\% of the time. Thus, these aspects of melody are easily controllable.

Previous work \cite{dai2020} has shown that pitch and rhythm distributions are related to different levels of long-term structure. We confirmed that our generation exhibits similar structure-related distributions to that of the POP909 dataset. 
For example, the probability of a generated tonic at end of a phrase is 48.28\%, and at the end of a section is 87.63\%, while in the training data the probabilities are 49.01\% (phrase-end) and 86.57\% (section-end).

\subsection{Subjective Evaluation}
\subsubsection{Design of the listening evaluation}
We conducted a listening test to evaluate the generated songs. To avoid listening fatigue, we presented sections lasting about 1 minute and containing at least 3 phrases.
We randomly selected 6 sections from different songs in the validation set as seeds and then generated melodies based on conditions 1--6 presented in Table \ref{tab:evaluation}. To render audio, each melody is mixed with the original chords played as simple block triads via a piano synthesizer.  For each section and each condition, we generated at least 2 versions, with 105 generated sections in total.

In each rating session, a listener first enters information about their music background and then provides ratings for six pairs of songs. Each pair is generated from the same input seed song using different generation conditions (see Table \ref{tab:evaluation}). For each pair, the listener answers: (1) whether they heard the songs before the survey (yes or no); (2) how much they like the melody of the two songs (integer from 1 to 5); and (3) how similar are the two songs' melodies (integer from 1 to 5). We also embedded one validation test in which a human-composed song and a randomized song are provided to help filter out careless ratings.

\subsubsection{Results and discussion}
We distributed the survey on Chinese social media and collected 274 listener reports. We removed invalid answers following the validation test and a few other criteria.
We ended up with 1212 complete pairs of ratings from 196 unique listeners. The demographics information about the listeners are as follows: \\
\hspace*{0.2in}\textit{Gender} male: 120, female: 75, other: 1;\\
\hspace*{0.2in}\textit{Age distribution} 0-10: 0, 11-20: 17, 21-30: 149,  31-40: 28,  41-50: 0, 51-60: 2, $>$60: 0;\\
\hspace*{0.2in}\textit{Music proficiency levels} lowest (listen to music $<$ 1 hour/week): 16, low (listen to music 1--15 hours/week): 62, medium (listen to music $>$ 15 hours/week): 21, high (studied music for 1--5 years): 52, expert ($>$ 5 years of music practice): 44;\\
\hspace*{0.2in}\textit{Nationality} Chinese: 180, Others: 16 (note that the POP909 dataset is primarily Chinese pop songs, and listeners who are more familiar with this style are likely to be more reliable and discriminating raters.)

\begin{figure}[tb]
 \centerline{
 \includegraphics[width=0.95\columnwidth]{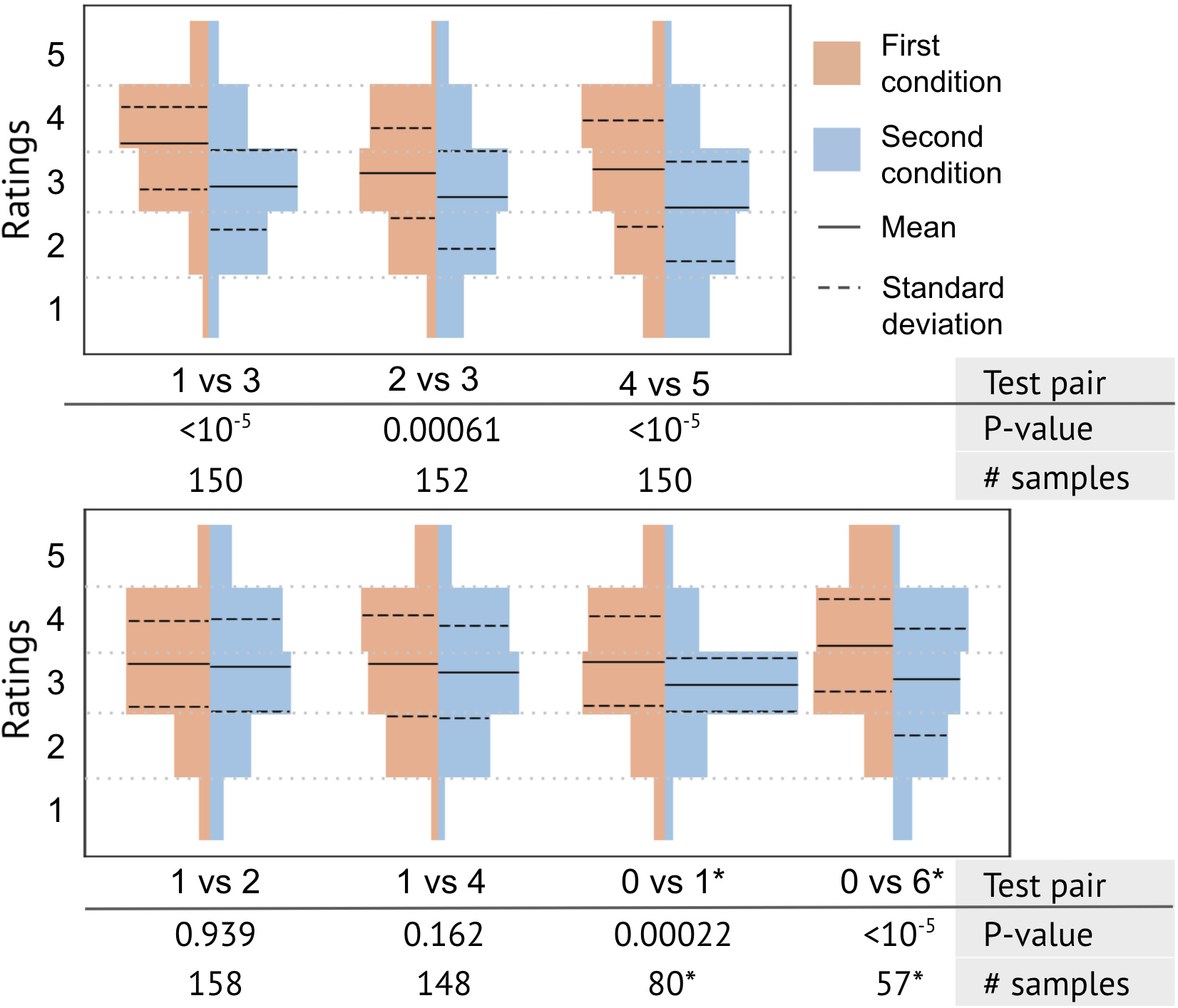}}
 \vspace{-0.1in}
 \caption{Rating distribution comparison for each paired groups. *For conditions 0 vs 1 and 0 vs 6, we removed the cases where the listeners indicated having heard the song before.}
 \label{fig:rating}
 \vspace{-1em}
\end{figure}

Figure~\ref{fig:rating} shows a distribution of ratings for the seven paired conditions in Table \ref{tab:evaluation}. In each pair, we show two bar plots with mean and standard deviation overlaid: the left half shows the distribution of ratings in the first condition and the right half shows those in the second condition. The first three pairs compare music generation with and without a music framework as an intermediate representation. The first two pairs at the bottom compare music with an automatically generated basic melody and rhythm to music using the basic melody and rhythm from a human-composed song. The last two pairs show the ratings of our method compared to music in the POP909 dataset. We also conducted a paired T-test to check the significance against the hypothesis that the first condition is not preferred over the second condition, shown under the distribution plot. 

\begin{figure}[tb]
 \centerline{
 \includegraphics[width=0.95\columnwidth]{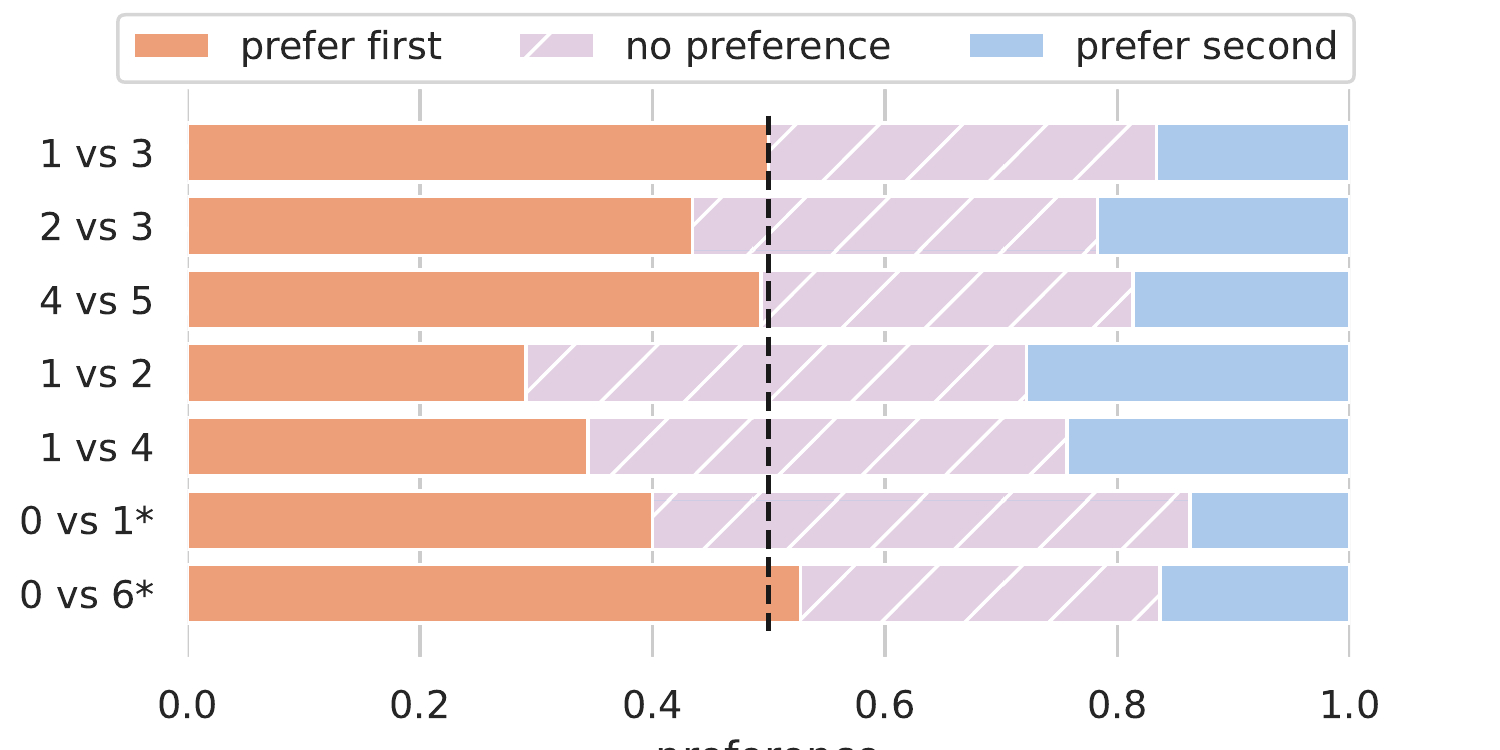}}
 \vspace{-0.1in}
 \caption{Preference distribution for each paired groups. *For conditions 0 vs 1 and 0 vs 6, we removed the cases where the listeners indicated having heard the song before.}
 \label{fig:pref}
 \vspace{-1em}
\end{figure}

In addition, we derived listener preference based on the relative ratings, summarized in Figure~\ref{fig:pref}. This visualization provides a different view from ratings as it shows how frequently one condition is preferred over the other or there is no preference (equal ratings). Based on these two plots, we point out the following observations:

\vspace{0.2em}
\begin{itemize}[leftmargin=*,noitemsep, nolistsep]
    \item Basic melody and basic rhythm form improve the quality of generated melody. Indicated by low p-values and strong preference in ``1 vs 3'', ``2 vs 3'' and ``4 vs 5,'' generating basic melody and basic rhythm before melody generation has higher ratings than generating melody without these music framework representations. 
    \item Melody generation conditioned on generated basic melody and basic rhythm has similar ratings to melody generated from human-composed music's basic melody and basic rhythm form. This observation can be derived from similar distribution and near random preference distribution in ``1 vs 2'' and ``1 vs 4,'' indicating that preference for the generated basic melody and rhythm form are close to those of music in our dataset. 
    \item Although both distribution tests suggest that human-composed music has higher ratings than generated music in test pairs ``0 vs 1'' and ``0 vs 6'' (and this is statistically significant), the preference test suggests that around half of the time the generated music is as good as or better than human-composed music, indicating the usability of the \textit{MusicFrameworks} approach. 
\end{itemize}
\vspace{0.2em}

\begin{table}
 \begin{center}
 \begin{tabular}{|l|l|l|l|}
  \hline
   & R.Melody & Basic Melody & Rhythm                              \\ 
\hline
0 & copy        & copy         & copy                                \\ 
\hline
1 & gen        & copy         & copy                                \\ 
\hline
2 & gen        & gen    & copy                                \\ 
\hline
3 & gen        & without      & copy                                \\ 
\hline
4 & gen        & copy         & gen with BRF     \\ 
\hline
5 & gen       & copy         & gen without BRF  \\ 
\hline
6 & gen       & gen     & gen with BRF    \\
\hline
 \end{tabular}
\end{center}
\vspace{-1em}
\caption{Seven evaluation conditions. Group 0 is human-composed. R.Melody: realized melody;
 gen: generated from our system; BRF: Basic Rhythm Form; copy: directly copying that part from the human-composed song; without: not using music frameworks. }
 \label{tab:evaluation}
\end{table}

To understand the gap between our generated music and human-composed music, we look into the comments written by listeners and summarize our findings below: 

\vspace{0.2em}

\begin{itemize}[leftmargin=*,noitemsep,nolistsep]
    \item Since sampling is used in the generative process, there is a non-zero chance that a poor note choice may be generated. Though this does not affect the posterior probability significantly, it degrades the subjective rating. Repeated notes also have an adverse effect on musicality with a lesser influence on posterior probability.
    \item \textit{MusicFrameworks} uses basic melody and rhythm form to control long-term dependency, i.e., phrases that are repetitions or imitations share the same or similar music framework; however, the generated melody has a chance to sound different due to the sampling process. A human listener can distinguish a human-composed song from a machine-generated song by listening for exact repetition.  
    \item Basic melody provides more benefit for longer phrases. For short phrases (4-6 bars), generating melodies from scratch is competitive with generating via basic melody. 
    \item The human-composed songs used in this study are from the most popular ones in Chinese pop history. Even though raters may think they do not recognize the song, there is a chance that they have heard it. A large portion of the comments suggest that a lot of the test music sounds great and it was an enjoyable experience working on these surveys. However, some listeners point out that concentrating on relatively long excerpts was not a natural listening experience. 
\end{itemize}

\vspace{0.2em}

\vspace{-0.5em}
\section{Conclusion}\label{sec:conclusion}

\textit{MusicFrameworks} is a deep melody generation system using hierarchical music structure representations to enable a multi-level generative process. The key idea is to adopt an abstract representation, \textit{music frameworks}, including long-term repetitive structures, phrase-level \textit{basic melodies} and \textit{basic rhythm forms}. We introduced analysis algorithms to obtain music frameworks from songs. We created a neural network that generates basic melody and additional networks to generate melodies. We also designed musical features and encodings to better introduce musical inductive bias into deep learning models. 

Both objective and subjective evaluations show the importance of having music frameworks. About 50\% of the generated songs are rated as good as or better than human-composed songs. Another important feature of the \textit{MusicFrameworks} approach is \textit{controllability} through manipulation of music frameworks, which can be freely edited and combined to guide compositions.

In the future, we hope to develop more intelligent ways to analyze music and music frameworks supporting a richer musical vocabulary, generation of harmony and polyphonic generation. We believe that hierachical and structured representations offer a way to capture and imitate musical style, offering interesting new research opportunities. 

\vspace{-0.5em}
\section{acknowledgments}
We would like to thank Dr. Stephen Neely for his insights on musical rhythm.

\bibliography{ISMIRtemplate}

\end{document}